\documentclass{llncs}

\usepackage[cmex10]{amsmath}
\usepackage{amsfonts,amssymb,graphicx}
\usepackage[mathscr]{eucal}
\usepackage{mathtools}
\usepackage[subnum]{cases}

\graphicspath{ {./img/} }
\DeclareGraphicsExtensions{.pdf}

\usepackage{cite}

\usepackage{caption}

\usepackage{url}

\usepackage{verbatim}

\usepackage{booktabs}
\usepackage{multirow} 

\usepackage{fixltx2e}

\begin{document}

\title{Resilience in Numerical Methods: A Position on Fault Models
and Methodologies}


\author{James Elliott\inst{1,2},
Mark Hoemmen\inst{1},
Frank Mueller\inst{2}
}

\institute{Sandia National Laboratories
\and
North Carolina State University
}

\maketitle



\begin{abstract}
  Future extreme-scale computer systems may expose silent data
  corruption (SDC) to applications, in order to save energy or
  increase performance.  However, resilience research struggles to
  come up with useful abstract programming models for reasoning about
  SDC.  Existing work randomly flips bits in running applications, but
  this only shows average-case behavior for a low-level, artificial
  hardware model.  Algorithm developers need to understand worst-case
  behavior with the higher-level data types they actually use, in
  order to make their algorithms more resilient.  Also, we know so
  little about how SDC may manifest in future hardware, that it seems
  premature to draw conclusions about the average case.  We argue
  instead that numerical algorithms can benefit from a \emph{numerical
    unreliability} fault model, where faults manifest as unbounded
  perturbations to floating-point data.  Algorithms can use
  inexpensive ``sanity'' checks that bound or exclude error in the
  results of computations.  Given a \emph{selective reliability}
  programming model that requires reliability only when and where
  needed, such checks can make algorithms reliable despite unbounded
  faults.  Sanity checks, and in general a healthy skepticism about
  the correctness of subroutines, are wise even if hardware is
  perfectly reliable.
\end{abstract}

\section{Introduction}

Much resilience research has focused on tolerating parallel process
failures through standard techniques like checkpoint/restart (C/R)
or process replication.
The monster in the closet \cite{geist11} is incorrect hardware
behavior which does not cause process failure.  Faults like incorrect
arithmetic or memory corruption may make the application produce
incorrect results or increase run time, with no indication from the
system that something went wrong.  We refer to this class of faults as
\emph{silent data corruption} (SDC).  If systems cannot correct these
faults before they affect applications' behavior, then the burden of
tolerating them shifts to algorithms: either to detect abnormal
behavior and correct it, or to ``absorb'' its effects while still
making progress towards the correct solution.
SDC's causes are poorly understood. Moreover, this type of fault is
rare enough that it is difficult to observe \cite{michalak12}.

Resilience approaches used in daily practice, such as C/R, are
attractive because they presume an abstract, minimal fault model.  C/R
assumes only that checkpoints contain correct state, and are stored
reliably to stable (shared, non-volatile) storage.  It does not care
whether the application failed due to a crashed node, network errors,
a power outage, or the job running out of allocation time.  This
abstract fault model lets C/R recover from many different kinds of
faults.

SDC, by definition, does not trigger system actions like a restart.
The silent error can manifest in several ways, such as performance
variation between parallel processes, convergence to a wrong solution,
or even an application crash sufficiently long after a checkpoint is
written that contains the tainted state.
%
%
Given the success of C/R's abstract fault model, why then has soft
error research focused so heavily on a low-level, fine-grained ``bit
flip'' fault model?  Bits may go wrong if an error is introduced, but
this does not help us design algorithms which work at a much higher
abstraction layer than bits.  Applications care about data types like
floating-point values and integers, not about the bits which compose
them.

We argue that resilient numerical methods should be designed around an
abstract fault model of \emph{numerical unreliability}, in much the
same way C/R is designed around an abstract model of system
unreliability.  We present a case for a radically different research
methodology that merges numerical analysis with systems fault
tolerance, and provides algorithm developers with programming models
they can use to ensure correctness despite SDC.  We solicit this
community specifically, because this challenge requires researchers
that are comfortable bridging mathematics and computer science.


\section{Bit Flips for Algorithm Analysis?}

All prior work in fault-tolerant numerical methods, including some of
ours, has presumed a \emph{bit flip} model of hardware faults. 
That is, faults occur randomly (e.g., via a Poisson process), and manifest
as one or more bits of one or more words changing values (``flipping'').
That word could store data of any type, including floating-point
values, integer indices, pointers, or even instructions.  A fault
itself does not immediately cause the affected process to crash,
except through any resulting changes to application behavior, such as
a segmentation violation signal raised due to an illegal memory access
caused by a corrupted pointer.

This model is seductive, because it lets researchers apply a ``computer
science'' approach to numerical algorithms.  In particular, it allows
\emph{stochastic sampling}.  That is, for a given problem, one either
randomly
\cite{bronevetsky08,sloan12,shantharam11,shantharam12,fiala12-2,%
howle10,du12,chen2013online,davies11,davies13}
or exhaustively \cite{elliott13,elliott13representation} flips bits
and observes the effects on running codes.  This makes for reasonable
papers: One picks some numerical algorithm (e.g., an iterative linear
solver), a set of test problems, and fires them off with some random
fault injection in the background.
The authors may also engineer a technique that detects and corrects
the faults being injected.

We argue against this approach.
First, we have no idea if real hardware faults manifest in this way,
either in current or future computers.  Second, stochastic sampling
tends to reveal average-case behavior, but we are most interested in
worst-case behavior.  
%
%
Third, the bit flip model does not reflect what
algorithms actually need to know, in order to increase their
resiliency. That is, simply pointing out that a bit flip can cause the
algorithm to stagnate, produce a comically wrong solution, or, if
lucky, get the right answer does not enable us to design  algorithms
any wiser than we did prior to the study. What this type of research
does accomplish, is that it shows that certain fault models can be
addressed via specific engineering approaches, e.g., checksums.

%
\subsection{Bit Flip Model May Not Reflect Actual Hardware Behavior}

Bit flips can manifest all
sorts of problems, from corrupting arithmetic results or storage to
changing the instruction stream.  There are too many ways in which
things could go wrong, so it's not clear where to start predicting
behavior.  For example, what happens if data in a cache are corrupted?
 It depends on whether the algorithm reads or writes the corrupted
data, as well as the cache eviction policy.  It is also not clear
whether possible \emph{future} hardware which lets faults through to
applications will behave according to our models.  We barely even know
how to program future fault-free hardware.

Future architectures may need to expose more unreliability to save
energy or improve performance. The question is what level of
unreliability a numerical technique can handle. We can make progress
towards this answer by focusing on research on algorithms that bound
error, rather than attempting to detect and correct all errors.  We
explain this in the following sections.

\subsection{Worst-Case Behavior: Adversarial vs.\ Practitioner}

Stochastic sampling is a natural tool to use given the bit flip
model.  Random sampling, in itself, is not bad, but as a
means to justify the correctness of a numerical method it is
inadequate. Numerical methods typically have \emph{proven} behavior
and correctness.  If operations can be unreliable, then we need to
identify the bounds in which a resilient algorithm is reliable. That
is, we should understand the smallest and largest errors we may
commit.  Relying on sampling alone leaves us prone to a practitioner
design pattern, where things are fixed only when someone (or a sample)
identifies there is a problem.  We feel a more adversarial approach is
required, and this approach fits naturally with a \emph{bounded error}
design methodology.  For a given numerical kernel, we wish to know the
worst-case error that can be committed, and with this knowledge, we
can employ pure and applied mathematicians to aid us in designing
methods that can tolerate such error bounds.

When developing an algorithm, we cannot ignore the extreme cases,
because if we do so, we have unstated assumptions about the way in
which the algorithm can be used. For numerical methods, unstated
assumptions make the method nearly worthless, given that we can never
anticipate what combinations of data the user will throw at the
algorithm. For this reason we advocate moving from a bit flip
model to a more abstract model that evaluates algorithms based on
their ability to absorb unexpected numerical variability.

\subsection{Large Perturbations and Boundedness}
The ``random bit flip'' fault model does not reflect what algorithms
actually need to know.  Algorithm developers do not care whether a
network packet dumped garbage into our reduction or a cosmic ray
blasted 20 entries in an array.  All of these events can be modeled as
numeric perturbations in the algorithm.  Furthermore, we can \emph{bound}
these errors by detecting their effects.  Then we can use numerical
approaches that can tolerate ``large'' errors, where ``large'' means ``much
larger than rounding error, but not large enough to detect.''

Our training leads us to restrict our consideration to \emph{numerical
  algorithms}, that is, approximations for the solution of continuous
problems using floating-point numbers.  Other authors have studied
ways to make discrete algorithms (like sorting) more fault tolerant,
by relaxing them into continuous problems \cite{sloan2011numerical}.
Mathematics has a long history of analyzing
the effects of perturbations to the input or intermediate results of numerical
algorithms.  Usually, those perturbations are small and represent the
effects of rounding error or the limited accuracy of experimental
data.  However, analysis has shown that some algorithms can tolerate
errors of size comparable to the input data (see e.g., inexact
Krylov \cite{simoncini2003theory}).
We are mainly interested in worst-case behavior, so we can exclude
small errors, because those are already covered by rounding error
analysis.

Under our abstract model of numerical unreliability and in conjunction
with bounds, then we know the worst-case faults lie within our bounds. 
This approach gives us a clear research direction that we feel will
prove extremely useful.  By analyzing algorithm's behavior to
``large'' perturbations within the algorithms theoretical bounds, we
can use analysis and experimentation to identify the worst-case
faults.

\subsection{What About Metadata?}

One might question whether it suffices just to consider floating-point
arithmetic and storage.  Numerical algorithms do not only have
\emph{data}, they also have \emph{metadata}: Pointers, indices,
program counters, and even the instructions themselves.  We consider
data to be the state required \emph{theoretically} by the numerical
method, e.g., a Krylov subspace $\mathcal{K}$, an input matrix $A$, a
right-hand side vector.  The metadata is the information required to
\emph{implement} the method, like integer dimensions $n$, loop
counters $i$, and sparse matrix indices.  Some metadata may occupy
space proportional to the data.  For example, with many sparse matrix
storage formats, indices take space proportional to that of the
matrix's values.  Does it make sense to consider data corruption,
without including the metadata?

We should always apply research effort in the most extensible way
possible. A numerical algorithm like the method of conjugate gradients
can be implemented in many ways, but its theoretical foundation will
remain the same. That is to say, the data as well as certain invariant
properties can be assumed \emph{a priori}, and we can harden the
algorithms by devising mechanisms to assert that these theoretical
principals remain true. The metadata required to implement the
algorithm can change drastically based on who implements the
algorithm, and the language or libraries chosen to do so. We will
expand this thought in the subsequent section.

There are three possible effects of metadata corruption.  First, it
could manifest as a floating-point data fault.  For example, a
corrupted array index would result in a read of the wrong value.
Second, it could crash the process, for example due to a segmentation
violation or an invalid instruction.  Third, it could be possible for
metadata corruption to let the process keep running, but put it in an
undefined, unpredictable state.\footnote{See the ``nasal demons''
  entry of \cite{raymond1996new}.}  Experience suggests that the third
option is exceptionally unlikely.  Furthermore, the second case
reduces to the known problem of process failure.  An entire genre of
research and practical software exists to handle this case.  Sometimes
the first case gets turned into the second, for example ECC memory
when it detects an error it cannot correct.

We can only argue about what to do for data faults, not whether
exposing data faults is a good idea.  If systems do expose faults to
applications, the metadata issue will arise.

\section{Numerical Unreliability and Skeptical Programming}

If we assume that operations are inherently unreliable, then we should
anticipate events such as $2+2 = 0$.  One approach to tolerate
numerical unreliability, is to detect and correct all errors. This is
what traditional Application-Based Fault Tolerance (ABFT)
\cite{huang1984} has done. ABFT methods propagate checksums
throughout an algorithm and use these checksums to assert that computations are correct.  This
approach is daunting, as the burden to detect and correct \emph{all}
numerical errors is difficult and an open area for research, even for
well understood numerical methods such as LU decomposition
\cite{davies13}. We advocate a different strategy.

Traditional ABFT attempts to preserve the illusion of an
always reliable machine.  Instead, we favor an approach more
compatible with numerical analysis.  First, we bound the error that
faults can introduce. Second, we identify methods that can tolerate
the largest error possible. This strategy is based entirely on the
algorithm theory and the data itself. That is, given specific inputs,
we can bound large portions of an algorithm using standard norm bounds
and inner product bounds.  These bounds enforce that errors committed
in intermediate computations do not exceed the theoretical limit
imposed by the algorithm in conjunction with the data provided. We
demonstrate this approach in \cite{elliott13gmres} and a similar
approach is used in \cite{vanDam13}.

Because operations are unreliable (numerical unreliability), the
bounds allow us to be \emph{skeptical} of key values. We use a bound
on orthogonal projections in \cite{elliott13gmres}, while Van Dam et
al.\ use a bound on a crucial inner product \cite{vanDam13}. These
bounds work as \emph{filters} rather than error detectors. We make no promise to
detect and correct all errors, we merely promise \emph{bounded error}.
We refer to this approach as \emph{Skeptical Programming}.

Numerical research provides a continuous stream of results that could
be used in our Skeptical Programming strategy. The key factor is that
these bounds be 1) cheap to evaluate, and 2) be determined
\emph{before} the algorithm is run.  We prefer to determine these
bounds \emph{a priori} because numerical unreliability may affect any
bound computed inside the algorithm. That is, if we allow a bound to
depend on unreliable computation, then the bound becomes unreliable as
well. We may have to relax (2) in some cases, but we desire (1) to be
true always. We clarify this reasoning by introducing \emph{Selective
  Reliability}.

\section{Selective Reliability}

Hoemmen and Heroux proposed a fault tolerance approach based on the
concept of isolating numerical operations that \emph{must} be
reliable, from those where reliability can be relaxed
\cite{bridges12fault}.  They use this to develop a programming model
that ``sandboxes'' unreliable computations, and promises reliability
on specific computations. A realization of Selective Reliability is
the two-level iterative solver FT-GMRES.  FT-GMRES preconditions
Flexible GMRES (FGMRES) by GMRES, possibly with its own
preconditioner. The outer FGMRES iteration is identified as needing
reliability, while the inner GMRES (and its preconditioner) is marked
as suitable for unreliability. In this setup, the outer iteration
absorbs the error introduced from numerical unreliability, while still
making progress towards the correct solution.

Skeptical Programming enhances Selective Reliability by bounding the
error that the inner (nested) GMRES solver can introduce. This
approach allows the inner solver to run without expensive
fault tolerance checks, such as frequently re-checking orthogonality
or computing the explicit residual.

The key is that Selective Reliability does not describe what can be
unreliable.  Instead, it \emph{only} declares what \emph{must} be
reliable. This approach enables phenomenal flexibility. For example, a
reliable FGMRES outer solver, can wrap complicated (black box)
preconditioners, while still promising that if a solution is obtained
it will be correct. This approach is in stark contrast to current
trends in fault tolerant algorithms, where algorithm developers are
attempting to robustify every numerical method to handle faults.

\section{Conclusions}


The resilience community has almost no idea how SDC will manifest in
future computers.  We just know that it \emph{may} show up.  Thus, we
aim to suggest models and best practices for algorithm developers,
that assume as little as possible about how faults appear.  Thinking
of SDC as unbounded perturbations to floating-point data, rather than
bit flips, describes it in a way useful to numerical analysts.  

An easy way to start using this model, is to introduce inexpensive
``sanity checks'' that help bound or exclude incorrect results.  These
checks are never a bad idea, because they can protect code against
many conditions other than SDC.  These include violated assumptions
about the input (e.g., that the matrix is nonsingular), rounding error
that unexpectedly violates physical constraints such as energy
conservation, and software bugs.  Algorithm experts are the right
people to design these checks.  They should favor checks that can
reduce error while bounding or measuring it.  This includes iterative
refinement for solving linear systems \cite{demmel2006extra}, along
with other inner-outer iterations like FT-GMRES.  Algorithms with
bounded error, such as regularized least squares instead of LU with
partial pivoting, make good building blocks for constructing more
resilient applications.

Combining these recommendations with a selective reliability
programming model will let applications \emph{prove} correctness.
Even without selective reliability, implementing these recommendations
should increase their resilience to SDC, as well as to other kinds of
events that applications already encounter on today's hardware.

\section*{Acknowledgments}


This work was supported partly by the RX-Solvers grant from the
Advanced Scientific Computing Research program of the U.S. Department
of Energy's (DOE) Office of Science, and partly by the Consortium for
Advanced Simulation of Light Water Reactors under U.S. DOE Contract
No. DE-AC05-00OR22725.  Sandia National Laboratories is a multiprogram
laboratory managed and operated by Sandia Corporation, a wholly owned
subsidiary of Lockheed Martin Corporation, for the U.S.\ DOE's
National Nuclear Security Administration under Contract
DE-AC04-94AL85000.

\bibliographystyle{plain}
\bibliography{paper}
\end{document}